\documentclass[]{elsarticle}
\usepackage{amssymb}
\usepackage{amsfonts}
\usepackage{amsmath}
\usepackage{graphicx}

\begin{document}
\title{How much quantum noise of amplifiers is detrimental to entanglement}
\author[rt]{G. S. Agarwal}
\ead{girish.agarwal@okstate.edu}
\author[fi]{S. Chaturvedi}
\ead{scsp@uohyd.ernet.in}
\address[rt]{Department of Physics, Oklahoma State University,
Stillwater, OK 74078, USA}

\address[fi]{School of Physics, University of Hyderabad, Hyderabad
500046, India}

\begin{abstract}
We analyze the effect of the quantum noise of an  amplifier on the
entanglement properties of an input state. We consider both phase
insensitive and phase sensitive amplification and specialize to
Gaussian states for which entanglement measures are well
developed. In the case of phase insensitive amplification in which
both the modes are symmetrically amplified, we find that the
entanglement in the output state vanishes if the intensity gain
exceeds a limiting value $2/(1+\exp[-E_N])$ where $E_N$ is the
logarithmic negativity of the input state which quantifies the
initial entanglement between the two modes. The entanglement
between the two modes at the output is found to be more robust if
only one mode is amplified.
\end{abstract}

\maketitle

Modifications of quantum features of light propagating through
absorbing and amplifying media has been a subject of considerable
importance and hence of intense activity  \cite{1}--\cite{12}. For
obvious practical reasons, the process of amplification has been
of special interest and for the case of a single mode, using
specific mathematical models for amplification, various authors
have derived limits on the amplifier gain beyond which the noise,
intrinsic to the process of amplification, destroys special non
classical features such as squeezing and sub-Poissonian
statistics. Similar limits, for the case when the output ceases to
have any non classical properties at all, have also been obtained.
Motivated by these studies, our aim here is to answer similar
questions in the context of quantum entanglement which has now
come to be universally recognized as an indispensable resource for
quantum information processing. To be specific, we consider the
effect of amplification on the two mode squeezed vacuum state
which falls in the category of much studied entangled Gaussian
states which occupy a privileged position in quantum information
processing through continuous variable systems \cite{13}. Not only
have they been put to use in experimental demonstrations of
teleportation \cite{14} and quantum cryptography, they have also
been found to be amenable to resolution of questions concerning
their separability \cite{15}--\cite{17} and distillability
\cite{18}.

In this letter we consider two kinds of amplifiers- phase
insensitive and phase sensitive. The two amplification processes
are quite distinct in character in so far as their entanglement
properties are concerned. While the phase insensitive amplifier
performs  local operations on the two modes, albeit non unitary,
phase sensitive amplification operates quite differently. Our key
results on the critical values of the gain are given by eqs.
$(\ref{f})$ and $(\ref{h})$. For phase sensitive amplifier we give
a condition on the phase mismatch that can be tolerated before the
entanglement in the output starts deteriorating. Interestingly
enough, Josse et al \cite{11} have demonstrated a linear optical
implementation of optical amplification process and some of the
ideas of this paper can be tested by methods of linear optics. We
consider phase
insensitive amplification  first.\\
\noindent {\it I. Phase Insensitive Amplifier}:~ We model this
type of amplifier in the standard way as a bath consisting of $N$
two level atoms of which $N_1$ are in the excited state and $N_2$
in the ground state with $N_1>N_2$. Under the assumptions that
atomic transitions have a large width and that the bath is
maintained in a steady state, the time evolution of the density
operator $\rho$ for a single mode of radiation field on resonance
with the atomic transition is described, in the interaction
picture,  by the master equation
\begin{eqnarray}
\frac{\partial\rho}{\partial t}&=&-\kappa
N_1(aa^\dagger\rho-2a^\dagger\rho a+\rho  aa^\dagger)\nonumber\\
&&-\kappa N_2(a^\dagger a \rho-2a\rho a^\dagger+\rho a^\dagger a),
\label{i}
\end{eqnarray}
where $a$ and $a^\dagger$ are the annihilation and creation
operators of the field mode. An optical realization of $(\ref{i})$
for $N_2=0$ is discussed in \cite{11}. An important feature of
this master equation is that if the initial density operator is a
Gaussian then its Gaussian character is preserved at later times.
In particular, an initial Gaussian Wigner distribution evolves
into a Gaussian Wigner distribution. The celebrated result of Hong
Friberg and Mandel \cite{3} states that both sub-Poissonian
statistics as well as the squeezing characteristics of any input
state (not necessarily Gaussian) survive if the gain $|G|^2$ where
$G\equiv \exp[\kappa t(N_1-N_2)]$ of the amplifier satisfies the
condition
\begin{equation}
|G|^2~<~\frac{2N_1}{N_1+N_2}~<~2.
\end{equation}

The evolution of the density operator under the master equation
above is entirely captured by the following intuitively appealing
Heisenberg like evolutions for the mode creation and annihilation
operators:
\begin{equation}
a(t)=Ga(0)+ c^\dagger,~  a^\dagger(t)=G^*a^\dagger(0)+ c^\dagger,
\label{a}
\end{equation}
where the averages of the noise operators $cc^\dagger$ and
$c^\dagger c$ are taken to be
\begin{equation}
<cc^\dagger>=(1+\eta)(|G|^2-1),~~<c^\dagger c>= \eta(|G^2|-1),
\label{b}
\end{equation}
with $\eta=N_2/(N_1-N_2)$. Hereafter we would take $(\ref{a})$ and
$(\ref{b})$, supplemented by the conditions $|G|^2>1$ and $\eta
\geq 0$ as the equations defining the action of a linear phase
insensitive quantum amplifier. Alternatively,  one may view these
equations as arising entirely from the requirements of
preservation of commutation relations supplemented by reasonable
phenomenological inputs. The condition that
$[a(t),a^\dagger(t)]=1$ ,given $[a(0),a^\dagger(0)]=1$, requires
that $[c,c^\dagger]=|G|^2-1$ and hence $<cc^\dagger>-<c^\dagger c>
=|G|^2-1$ and the noise correlations as in $(\ref{b})$ are
consistent with this equality. This reformulation of the master
equation above not only facilitates computations but also relates
directly to experimentally relevant quantities. It should be kept
in mind that the noise source is Gaussian in nature.

For continuous variable systems, entanglement measures are well
developed for Gaussian states and thus for our quantitative
studies we have chosen to examine the effects of quantum noise of
the amplifier on Gaussian entangled states.

In order to study the evolution of quantum entanglement of a
Gaussian state $\rho$  as it is amplified, in a manner as
prescribed above, we make use of the well known criteria for
entanglement in Gaussian states developed in \cite{15}--\cite{17}
where the covariance matrix $\sigma$ of the Wigner distribution,
\begin{eqnarray}
&&W(X)=\frac{e^{-(X-<\hat{X}>)\sigma^{-1}(X-<\hat{X}>)^T/2}}{(2\pi)^n\sqrt{{\rm
Det}(\sigma)}},\nonumber\\&&~~~~~~~~~~~~~~~~~~~~X\equiv(x_1,p_1,\cdots,x_n,p_n),
\end{eqnarray}
associated with the $n$ mode Gaussian state $\rho$ plays a key
role. The elements of covariance matrix $\sigma$, are given by
\begin{equation}
\sigma_{ij}=\frac{1}{2}<(\hat{X}_i\hat{X}_j+\hat{X}_j\hat{X}_i)>-<\hat{X}_i><\hat{X}_j>,
\end{equation}
where $\hat{X}\equiv
(\hat{x}_1,\hat{p}_1\cdots,\hat{x}_n,\hat{p}_n)$,
$\hat{x}_j=(a_j+a_j^\dagger)/\sqrt{2}$,
$\hat{p}_j=(a_j-a_j^\dagger)/\sqrt{2}i$, $a_j$ ($a_j^\dagger$)
denote the bosonic annihilation (creation) operators associated
with the $j^{\rm th}$ mode, and $<\cdot>\equiv {\rm Tr}[\cdot
\rho]$. By definition, the covariance matrix is a real positive
symmetric matrix and, by Williamson's theorem, is therefore
congruent by a symplectic transformation to a diagonal matrix. The
entries along the diagonal, the symplectic eigenvalues, can then
be used to constrain and characterize the  covariance matrices
associated with a quantum state \cite{21}. The idea that
violations of these constraints by the symplectic eigenvalues of
the `partially transposed' density operator $\tilde{\rho}$
corresponding to the Gaussian state $\rho$ can be viewed as a
signal for entanglement, has been ingeniously put to use in
\cite{15}--\cite{17} to arrive at the necessary and sufficient
conditions for separability of bipartite Gaussian states.

In the two mode case, the covariance matrix has the form
\begin{equation}
\sigma=\left(\begin{array}{cc}\alpha&\gamma\\
\gamma^T&\beta\end{array}\right).
\end{equation}
Given the covariance matrix for a two mode state $\rho$, the two
symplectic eigenvalues of the covariance matrix associated with
its `partial transpose' $\tilde{\rho}$ are explicitly given by ,
\begin{equation}
\tilde{\nu}_\pm= \sqrt{\frac{\tilde{\Delta}(\sigma)\pm
\sqrt{\tilde{\Delta}(\sigma)^2-4{\rm Det}(\sigma)}}{2}},
\end{equation}

\begin{figure}[!h]
\begin{center}
\includegraphics[scale = 0.3]{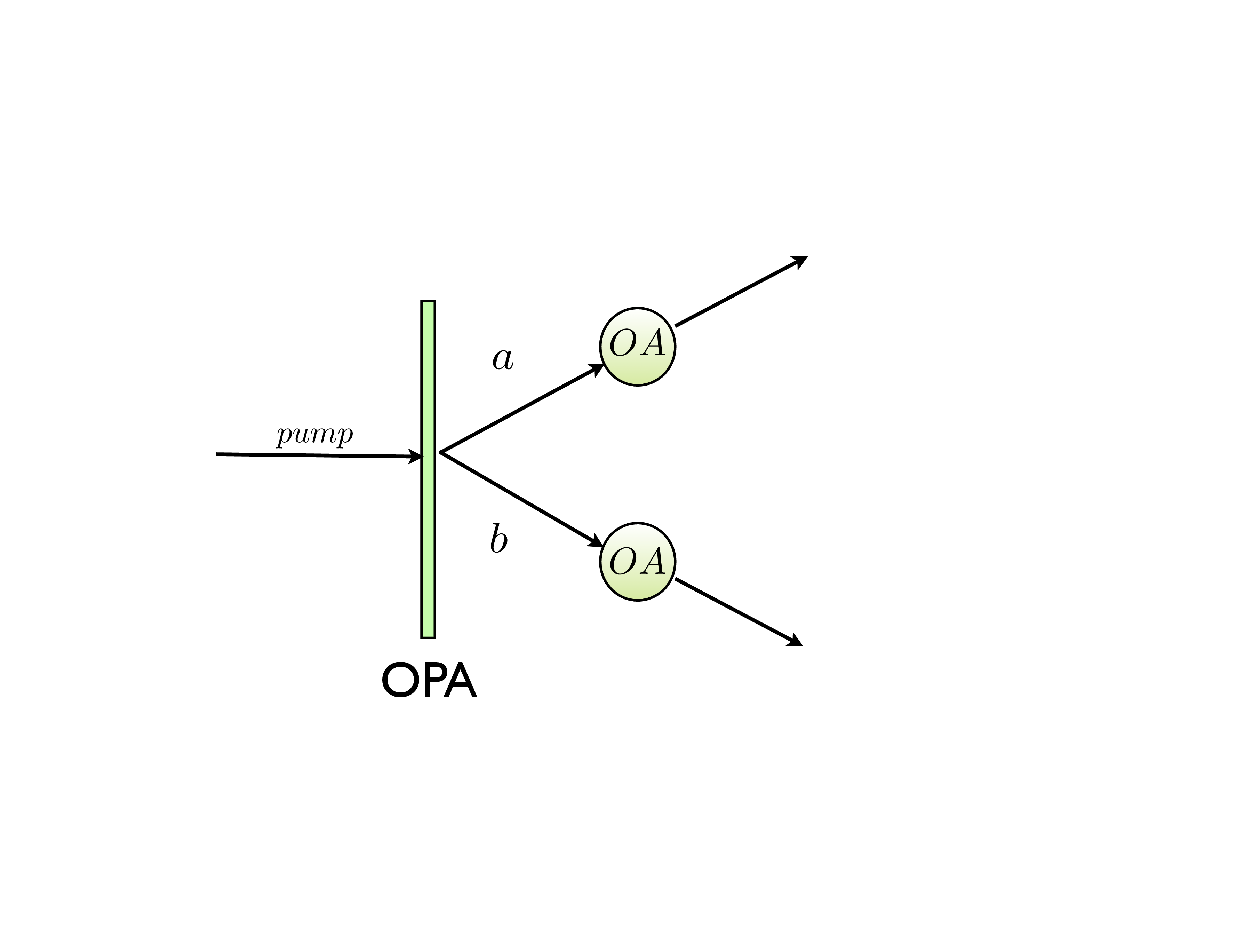}
\caption{Schematic diagram for the amplification of a two mode
entangled Gaussian state by a phase insensitive amplifier. The
optical parametric amplifier (OPA) produces a two mode squeezed
vacuum state of $a$ and $b$. In the symmetric case, both the optical
amplifiers (OA) are present. In the asymmetric case the OA from the
$b$ arm is removed. }
\end{center}
\end{figure}
\noindent where $\tilde{\Delta}(\sigma)={\rm Det}(\alpha)+{\rm
Det}(\beta)-2{\rm Det}(\gamma)$. If $\tilde{\nu}_<$ is taken to
denote the smaller of the two symplectic eigenvalues, then  the
necessary and sufficient conditions for $\rho$ to be an entangled
state can be expressed as
\begin{equation}
\tilde{\nu}_<<\frac{1}{2}.
\end{equation}
A possible quantitative measure of entanglement, discussed in
detail in \cite{p,22}, may be taken to be the logarithmic
negativity $E_N(\rho)$:
\begin{equation}
E_N(\rho)={\rm max}[0,-{\rm ln}(2\tilde{\nu}_<)].
\end{equation}
For the special case when $\sigma$ has the form
\begin{equation}
\sigma=\left(\begin{array}{cccc}A&0&B&C\\
0&A&C&-B\\B&C&A^\prime&0\\C&-B&0&A^\prime\end{array}\right),
\end{equation}
the two symplectic eigenvalues $\tilde{\nu}_\pm$ turn out to be
\begin{equation}
\tilde{\nu}_\pm \equiv
\frac{1}{2}[(A+A^\prime)\pm\sqrt{(A-A^\prime)^2+4(B^2+C^2)}].
\label{d}
\end{equation}
These expressions will be useful later.
For the two mode squeezed vacuum
\begin{eqnarray}
&&\rho=S(z)|0,0><0,0|S^\dagger(z),\nonumber\\
&& S(z)=\exp[za^\dagger b^\dagger-z^*a b],~ z=re^{i\theta},
\end{eqnarray}
which belongs to the class of Gaussian entangled states, the two
symplectic eignevalues are $\tilde{\nu}_\pm=e^{\pm 2r}/2$ and thus
the quantum entanglement in in this state is directly linked to
the squeezing parameter $r$ as the logarithmic negativity
$E_N=2r$. As $r\rightarrow 0$, $\tilde{\nu}_< \rightarrow 1/2$ and
the entanglement disappears. Using $(\ref{a})$ and $(\ref{b})$ we
next discuss the dynamics of
entanglement and consider two cases. \\
\noindent {\it A The Symmetric Case} Here we examine the situation
when both the modes are symmetrically amplified as shown
schematically in the Fig.1.

On making the replacements appropriate to this case
\begin{equation}
a\longrightarrow Ga+c^\dagger, b\longrightarrow Gb+d^\dagger,
\end{equation}
and using
\begin{eqnarray}
<cc^\dagger>=(1+\eta)(|G|^2-1),<c^\dagger c>= \eta(|G^2|-1),\label{e}\\
<dd^\dagger>=(1+\eta)(|G|^2-1),<d^\dagger d>= \eta(|G^2|-1),
\end{eqnarray}
with all other noise averages set equal to zero, one finds that
entries in $(\ref{d})$ are given by
\begin{eqnarray}
&&A=A^\prime =[|G|^2\cosh 2r+(1+2\eta)(|G|^2-1)]/2,\nonumber\\
&& B=[|G|^2\sinh 2r\cos\theta]/2, C=[|G|^2\sinh
2r\sin\theta]/2,\nonumber\\
\end{eqnarray}
and hence
\begin{equation}
\tilde{\nu}_<=[|G|^2(e^{-2r}+(1+2\eta))-(1+2\eta)]/2.
\end{equation}
Requiring that the output state remain an entangled state i.e.
$\tilde{\nu}_< <1/2$ then translates into the following condition
for the gain :
\begin{equation}
|G|^2<\left(\frac{2+2\eta}{1+2\eta+e^{-2r}}\right). \label{f}
\end{equation}
In the special case of a fully inverted amplifier,
$\eta\rightarrow 0$, and we have
\begin{equation}
|G|^2<\frac{2}{(1+e^{-2r})}=\frac{2}{(1+e^{-E_N})} , \label{g}
\end{equation}
where $E_N$ is the logarithmic negativity of the input state.
Further, for maximum entanglement in the initial state i.e.
$r\rightarrow \infty$, this becomes $|G|^2<2$. We note that Scheel {\it et al.}
\cite{Scheel1} using the concept of relative quantum entropy derived a result equivalent to $(\ref{g})$. Hillery and Zubairy \cite{Hillery} used the high gain limit and concluded that in this limit all entanglement would be lost. These papers however do not use the logarithmic negativity criteria. It is interesting to note that conditions similar to $(\ref{f})$ are found for the case of loss of entanglement \cite{16} by attenuators which have been extensively studies \cite{Scheel1,Scheel2,Huang}.
\begin{figure}[!h]
\begin{center}
\includegraphics[scale = 0.4]{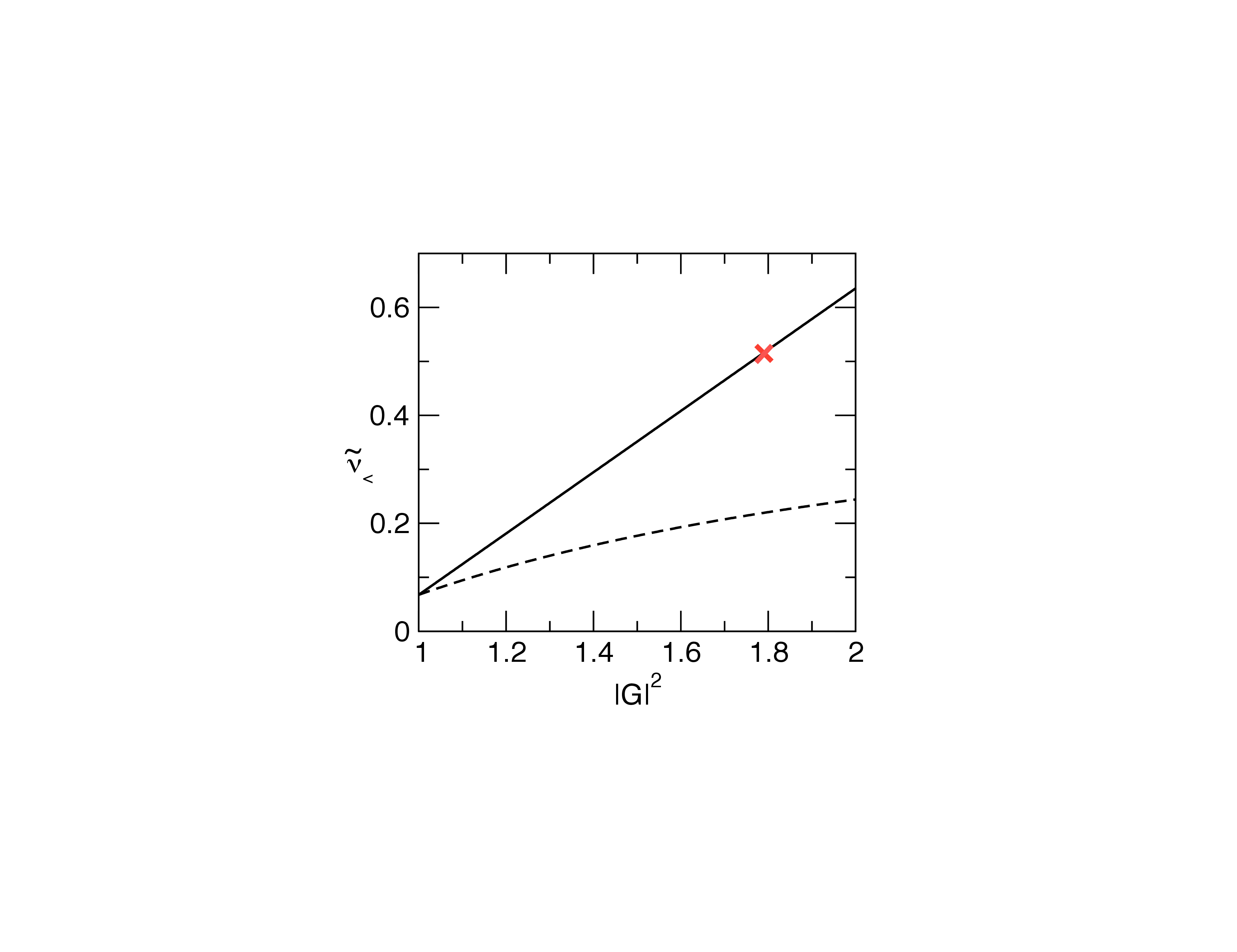}
\caption{Variation of the entanglement measure $\tilde{\nu}_< $ as a
function of the gain $|G|^2$ for the symmetric case (solid line)
and the asymmetric case (dashed line) for $r=1, \eta=0$. The cross
marks the critical value of the gain in the symmetric case beyond
which the entanglement in the output state vanishes.}
\end{center}
\end{figure}

\noindent {\it B The Asymmetric Case} Next we consider the case
when, say only the mode $a$ is amplified. On making the
replacements
\begin{equation}
a\longrightarrow Ga+c^\dagger, b\longrightarrow b,
\end{equation}
and  and using $(\ref{e})$ we obtain
\begin{eqnarray}
&&A=[|G|^2\cosh 2r +(1+2\eta)(|G|^2-1)]/2,\nonumber\\
&& A^\prime=[\cosh 2r]/2, B=[|G|\sinh 2r\cos\theta]/2,\nonumber\\
&&C=[|G|\sinh 2r\sin\theta]/2,
\end{eqnarray}
which give
\begin{eqnarray}
\tilde{\nu}_<&=&\frac{1}{4}[(|G^2|+1)\cosh 2r
+(1+2\eta)(|G|^2-1)\nonumber\\
&&-\sqrt{(|G|^2-1)^2(\cosh 2r+1+2\eta)^2+4|G|^2\sinh^2
2r}]\nonumber\\
\label{h}
\end{eqnarray}
In contrast to the symmetric case, here one finds $\tilde{\nu}_<$
is always less than $1/2$ for $\eta=0$ and therefore the entanglement survives
no matter how large the gain is. The variation of $\tilde{\nu}_<$
in the two cases as a function of the gain is shown in the Fig. 2.
The situation however is different if $\eta$ is non zero. In this case
there is a threshold value of gain beyond which the entanglement of the
initial state is lost. The threshold value of the gain depends on the value
of $\eta$. Clearly for larger $\eta$ the entanglement degrades faster. We show this behavior in the Fig 3. As mentioned in the introduction, our results like $(\ref{g})$ and
$(\ref{h})$ can be tested purely by means of linear optics
\cite{11}.
\begin{figure}[htp]
\begin{center}
\includegraphics[scale = 0.65]{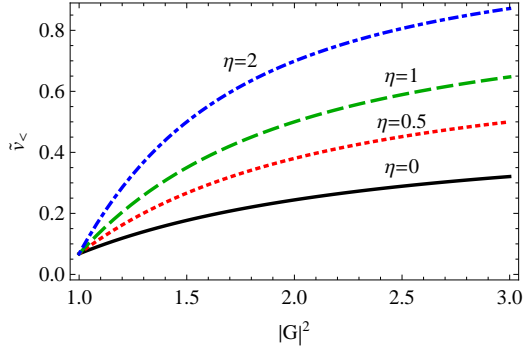}
\caption{Variation of the entanglement measure $\tilde{\nu}_< $ as a
function of the gain $|G|^2 $ for different $\eta$ and the squeezing parameter $r=1$. $\eta=0$(Solid
line), $\eta=0.5$(Dotted line), $\eta=1$(Dashed line),
$\eta=2$(Dotdashed line).}
\end{center}
\end{figure}
\noindent {\it II Phase Sensitive Amplification}: We next consider a
phase sensitive amplifier as shown schematically in the Fig. 4. The
net amplification in this case depends on the relative phase
$(\theta-\theta^\prime)$. Interesting interference effects in OPA
cavities due to pumping with squeezed light were predicted and
verified \cite{Agarwal}--\cite{Chen}.
\begin{figure}[!h]
\begin{center}
\includegraphics[scale = 0.3]{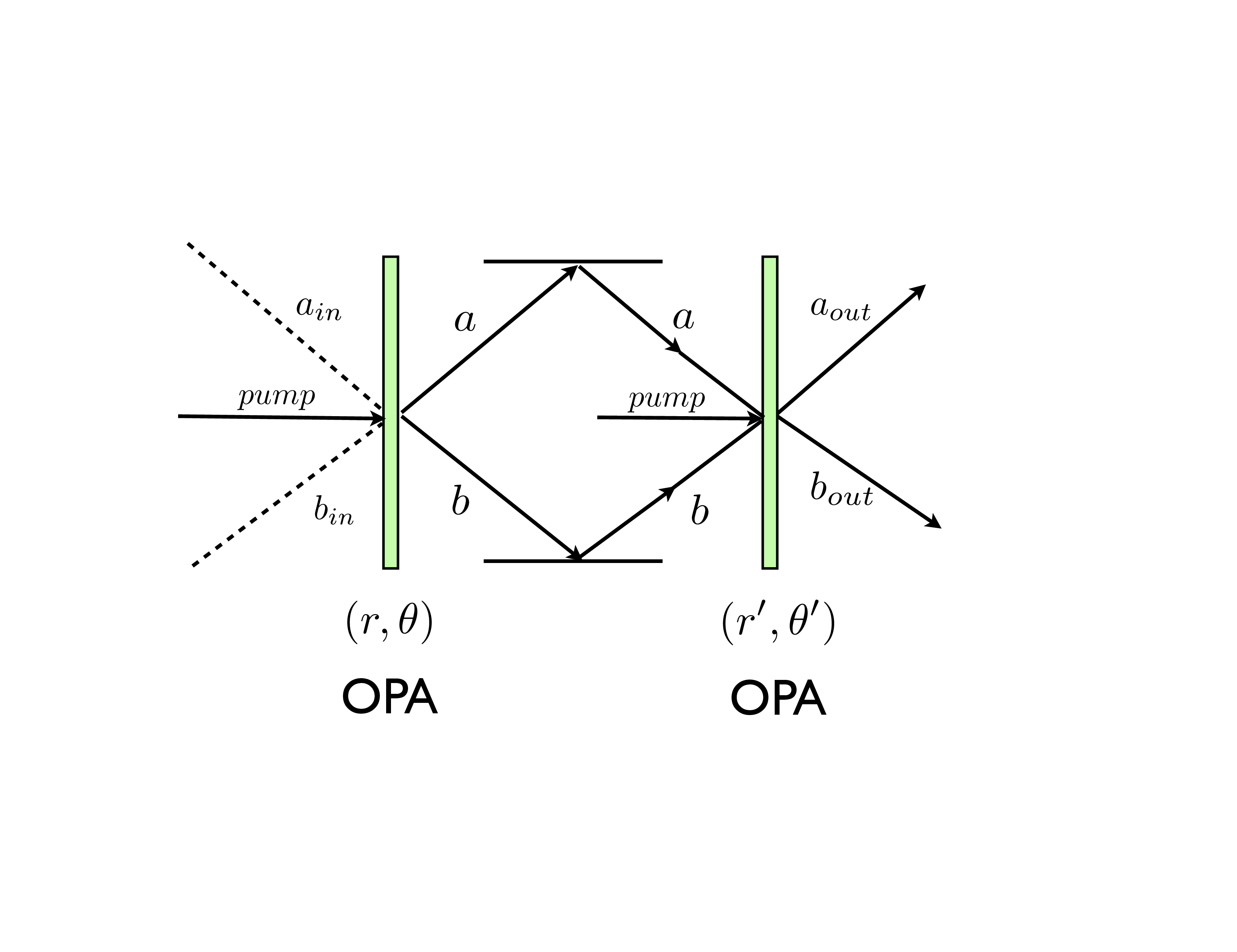} \caption{Schematic diagram of
a phase sensitive amplifier. The first OPA produces a squeezed
vacuum state of the modes $(a,b)$ and the second OPA acts as an
amplifier.}
\end{center}
\end{figure}

The action of this amplifier is to transform the input squeezed
vacuum $S(z)|0,0><0,0|S^\dagger(z)$ to the state
$S(z^{\prime})S(z)|0,0><0,0|S^\dagger(z)S^\dagger(z^{\prime})$.
Using the composition theorem for squeezing operators \cite{24},
\begin{eqnarray}
&&S(z_1)S(z_2)=S(z_3)e^{i(a^\dagger a+b^\dagger b+1/2)\Phi},~~z_i=r_ie^{i\theta_i},\nonumber\\
&&\zeta_3=\frac{\zeta_1+\zeta_2}{1+\zeta_1^*\zeta_2},
~~,\zeta_i=\tanh r_ie^{i\theta_i}\nonumber\\
&&\Phi=\frac{1}{2i}{\rm
ln}\left(\frac{1+\zeta_1\zeta_2}{1+\zeta_1^*\zeta_2}\right),
\end{eqnarray}
one finds that the output state is again a squeezed vacuum
$S(z^{\prime\prime})|0,0><0,0|S^\dagger(z^{\prime\prime})$, and
the squeezing parameter $r^{\prime\prime}$ is related to the
squeezing parameter $r^\prime$ of the input $(a,b)$ and
$r^{\prime\prime}$, that of the second OPA, as follows:
\begin{equation}
\cosh(2r^{\prime\prime})=\cosh(2r)\cosh(2r^\prime)+\sinh(2r)\sinh(2r^\prime)\cos\alpha,
\end{equation}
where $\alpha=|\theta-\theta^\prime|$. This relation sets a limit
on the phase difference $\alpha$ beyond which the squeezing
parameter for the output and hence its entanglement is less than
that of the input. This limiting value $\alpha_0$ of $\alpha$ is
given by
\begin{equation}
\cos\alpha_0 =\begin{array}{c}-\coth 2r\tanh  r^\prime~{\rm if}~ r\geq  r^\prime,\\
-\coth 2r^\prime\tanh r~{\rm if}~ r\leq r^\prime .\end{array}
\end{equation}
Thus, in order not to degrade quantum entanglement in the output,
 the phases $\theta$ and $\theta^\prime$ should be chosen such that
$0\leq |\theta-\theta^\prime| \leq \alpha_0 $.

To conclude, we derive the maximal allowable gain of the phase
insensitive amplifier for quantum entanglement to survive.
Interestingly enough, the entanglement is found to be more robust
if only one partner of the pair is amplified which is in agreement
with a recent observation \cite{lett}. Further, we give
explicit results for the growth of entanglement if phase sensitive
amplifier is used. Clearly extension of these ideas to non
Gaussian entangled states, like photon subtracted squeezed states
\cite{25} and NOON states \cite{26}, is a subject of further
study.
\\
\noindent {\bf Acknowledgements} One of us (GSA) has enjoyed
interesting conversations on entanglement with R. Boyd, J. H.
Eberly, P. Knight, M. Plenio and F. Illuminati. This work was
supported by NSF Grant no CCF-0829860.

\end{document}